\documentclass[aps,pra,twocolumn,showpacs,floatfix]{revtex4}
\usepackage{times,amsmath,amssymb,amsfonts,graphicx}
\newcommand{\ket}[1]{\vert #1 \rangle} %
\newcommand{\bra}[1]{\langle #1 \vert} %

\def\snr{{\rm SNR}}
\def\qsnr{{\rm QSNR}}

\def\opt{{\rm opt}}
\begin{document}
\title{Quantum probes for the spectral properties of a classical environment}
\author{Claudia Benedetti}
\affiliation{Dipartimento di Fisica, Universit\`a degli 
Studi di Milano, I-20133 Milano, Italy} %
\author{Fabrizio Buscemi}
\affiliation{ ARCES, Universit\`{a} di Bologna, Via Toffano 2/2, I-40125 Bologna,
Italy}
\author{Paolo Bordone}
\affiliation{Dipartimento di Scienze Fisiche, Informatiche e Matematiche, and
Centro S3, CNR-Istituto di nanoscienze,
Universit\`{a} di Modena e Reggio Emilia, via Campi 213/A,  Modena I-41125, Italy}
\author{Matteo G. A. Paris}
\affiliation{Dipartimento di Fisica, Universit\`a 
degli Studi di Milano, I-20133 Milano, Italy}
\affiliation{CNISM UdR Milano Statale, I-20133 Milano, Italy}
\date{\today}
\begin{abstract}
We address the use of simple quantum probes for the spectral
characterization of classical noisy environments.  In our scheme a qubit
interacts with a classical stochastic field describing environmental
noise and is then measured after a given interaction time in order to
estimate the characteristic parameters of the noise.  In particular, we
address estimation of the spectral parameters of two relevant kinds of
non-Gaussian noise:  random telegraph noise with Lorentzian spectrum and
colored noise with $1/f^{\alpha}$ spectrum. We analyze in details the
estimation precision achievable by quantum probes and prove that
population measurement on the qubit is optimal for noise estimation in
both cases. We also evaluate the optimal interaction times for the
quantum probe, i.e. the values maximizing the quantum Fisher information
(QFI) and the quantum signal-to-noise ratio. For random telegraph noise
the QFI is inversely proportional to the square of the switching rate,
meaning that the quantum signal-to-noise ratio is constant and thus the
switching rate may be uniformly estimated with the same precision in its
whole range of variation. For colored noise, the precision achievable in
the estimation of ``color'', i.e. of the exponent $\alpha$, strongly
depends on the structure of the environment, i.e. on the number of
fluctuators describing the classical environment.  For an environment
modeled by a single random fluctuator estimation is  more precise for
pink noise, i.e. for $\alpha=1$, whereas by increasing  the number of
fluctuators, the quantum signal-to-noise ratio has two local maxima,
with the largest one drifting towards $\alpha=2$, i.e. brown noise. 
\end{abstract}
\maketitle
In any communication channel or measurement scheme, the interaction of
the information carriers with the external environment introduces noise
in the system, thus degrading the overall performances.  The precise
characterization of the noise is thus a crucial ingredient for the
design of high-precision measurements and reliable communication
protocols. In many physical situations, the main source of noise is
associated to the fluctuations of  bistable quantities. In these cases,
a suitable description of the noise is given in terms of classical
stochastic processes \cite{band13, berman12,galp06}. In particular, in
the case of phase damping, i.e. pure dephasing, it has been shown that
the interaction of a quantum system with a quantum bath can be written
in terms of a random unitary evolution 
driven by a classical stochastic process \cite{joynt13,helm11}. 
\par
The characterization of classical noise is often performed by
collecting a series of measurements to estimate the autocorrelation
function and the spectral properties \cite{hur51,per03,barunik10,kend99,salomon13,
magd13}. This procedure is generally time
consuming and may require the control of a complex system. A question
thus arises on whether more effective techniques may be developed.  To
this purpose, we address the use of {\em quantum probes} to estimate the
parameters of classical noise. We assume to have a quantum system
interacting with the classical fluctuating field generating the noise 
and explore the performances of quantum measurements performed at a
fixed interaction time to extract information about the classical noise.
The power and implications of this idea are undeniable: the features of
a complex system may be determined by monitoring a small probe, which is
usually characterized by few and easily controllable degrees of freedom.
The simplest and
paradigmatic example of this situation is that of a qubit interacting with a
noisy environment.  After a given interaction time, which may be
suitably optimized, quantum-limited measurements on the qubit may be
used to characterize the environment, e.g. to estimate the parameters 
describing its noise spectrum. 
\par
In this paper we focus on the characterization of two classes of
classical noise: random telegraph noise (RTN) with Lorentzian spectrum
and the power-law $1/f^{\alpha}$ {\em colored} spectra arising
from the interaction with a collection of random bistable fluctuators.
Both RTN and colored noise are examples of classical non-Gaussian noise
occurring in several system of interest. Indeed, the microscopic models
underlying these kinds of noise have been
extensively analyzed in the literature
\cite{galperin,benIJ,bordonefnl,weissman,paladinorev,benedettipra,bus13}.  The
relevant parameters characterizing this kinds of noise are the switching
rate of the RTN and the exponent $\alpha$ in the case of power-law
spectra. Both quantities do not correspond to observables in a strict
sense and therefore we have to resort to indirect measurements performed
on the quantum probe to infer their value. In order to optimize this
inference procedure we employ tools from local quantum estimation
theory  \cite{hel76,mal93,brau94,brody99,paris09,esch12},
which have already been proved useful in the estimation
of static noise parameters \cite{hotta05,monras07,fuj01,ji08,dau06} and in several other
scenarios, as for example the estimation of quantum correlations
\cite{brida2010,brida2011,blandino2012,benedetti2013}, Gaussian
states \cite{mon10,pinel13,monar}, optical phase 
\cite{monras06,kac10,cable10,durk10,genoni2011,spagnolo2012},
critical systems \cite{zan08,inver08} and quantum thermometry \cite{brunelli2011}. 
In particular, we will optimize the initial preparation of the qubit and 
the interaction time in order to maximize the quantum Fisher information
and the quantum signal-to-noise ratio. Furthermore, 
we show that population measurement provides optimal inference
for both the noise models. 
\par
This paper is organized as follows: in Section \ref{sec1} we introduce
the physical dephasing model employed throughout the paper and describe
the main features of RTN and colored noise. In Section \ref{sec2}
we briefly review the main tools of quantum estimation theory, 
whereas in Section \ref{sec3} we present
our results on the precision achievable by quantum probes in the estimation of the spectral
properties of noisy random environments.  Section \ref{sec4} closes
the paper with concluding remarks.
\section{The Physical Model}\label{sec1}
In order to gain information about a complex environment, we analyze its
influence on the dynamics of a quantized information carrier. In the
simplest case this corresponds to a qubit interacting with a classical
stochastic field. Two different field spectra will be considered: the
Lorentzian spectrum generated by a random telegraph noise and the
$1/f^{\alpha}$ colored spectrum stemming from a collection of random
bistable fluctuators.  In both cases the noise induced by the classical
field is described by a non-Gaussian process, meaning that the sole
knowledge of the second-order statistics is not sufficient to fully
characterize the process.
\par
We focus attention on situations where the dominant process induced by 
the environmental noise is pure dephasing. This corresponds to have the 
quantum probe, a qubit, coupled to a classical field
in a given direction, say $x$. The Hamiltonian of the qubit
thus reads
\begin{equation}
 \mathcal{H}(t)=\epsilon\; \mathbb{I}+\nu\,c(t) \,\sigma_x,\label{hamiltonian}
\end{equation}
where $\epsilon$ is the energy of the qubit eigenstates assumed to be
degenerate, $c(t)$ is the stochastic non-Gaussian process, $\sigma_x$ is
the Pauli matrix, $\nu$ describes the coupling strength with the
environment and $\hbar$ was set to 1.  We also assume that the qubit is
initially prepared in a generic pure state
\begin{equation}
 \ket{\psi_0}=\cos\frac{\theta}{2}\,\ket{0}+e^{i\phi}\sin\frac{\theta}{2}\,
 \ket{1}.\label{psi0}
\end{equation}
Upon studying the dynamics of the qubit subject to noise we gain 
information about the spectral properties of the environment.
\subsection{Random telegraph noise}
Random telegraph noise describes the fluctuations induced by the 
interaction with a classical bistable fluctuator, i.e. a physical system which flips
between two given configurations with a fixed switching rate. RTN is characterized by an
exponential autocorrelation function and a Lorentzian
power spectrum. In mathematical terms, RTN corresponds to an 
interaction Hamiltonian as in Eq. \eqref{hamiltonian} with $c(t)$
flipping between the values $c(t)=\pm1$ at a switching rate $\xi$. 
\par
Hereafter we work with dimensionless quantities by scaling the 
time and the switching rate in unit of $\nu$. In particular,
we substitute $t\rightarrow\tau=\nu t$ and 
$ \xi\rightarrow\gamma=\xi/\nu$.\par
The density matrix of a qubit interacting with a RTN classical 
environment is obtained averaging the unitary evolved state 
over all possible temporal sequences of the stochastic process $c(t)$ 
\cite{benedetti1,paladino,abel}:
\begin{equation}
  \rho(\tau,\gamma,\theta,\phi)=\langle U(\tau)\rho_0U^{\dagger}(\tau)\rangle_{c(t)}\label{rrtn}
\end{equation}
where $U(\tau)=e^{-i\int_0^{\tau} {\mathcal H}(s)ds}$  is the evolution operator, 
$\rho_0=\ket{\psi_0}\bra{\psi_0}$ is the initial density matrix, and 
$\langle\dots\rangle_{c(t)}$ denotes average over the process.
The density matrix in the computational basis $\{|0\rangle,|1\rangle\}$  reads:
\begin{align}
\rho(\tau,\gamma,&\theta,\phi)=\nonumber\\
&\frac{1}{2}
(\cos\theta\, \mathbb{I}+ D(\tau,\gamma) \,\sigma_z+\nonumber\\
&\sin\theta\,\cos\phi\,\sigma_x-\sin\theta\,\sin\phi\,D(\tau,\gamma)\,\sigma_y)
\label{rtntp}
\end{align}
with 
\begin{align}
D(\tau,\gamma)
&\equiv \langle \text{exp}\left[-i\int_0^{\tau}c(s)\,ds\right]\rangle_{c(t)} \notag \\
&=e^{- \gamma \tau }\left(\cosh\delta\tau+
\frac{\gamma \sinh\delta\tau}{\delta}\right)
\end{align}
and where $\delta\equiv \delta(\gamma)=\sqrt{\gamma^2-4}$. For $\gamma<2$, $D(\tau,\gamma)$ is 
a damped oscillating function of time, while for $\gamma\geq2$  $D(\tau,\gamma)$ decays
monotonically in time. The first case is often referred to as slow RTN and corresponds
to a non-Markovian map \cite{bpm13}, while the
second is called fast RTN and leads to a Markovian dynamics.
\subsection{Colored noise}
A complex environment characterized by a noise spectrum
of the form $1/f^{\alpha}$ in a given frequency range 
$[\gamma_1,\gamma_2]$, correspond to a collection of one 
or more bistable fluctuators whose switching rates 
assume random values $f\in [\gamma_1,\gamma_2]$
according to the probability distribution
\begin{align}\label{distrib}
p_{\alpha}(\gamma)=\left\{\begin{array}{ll}
 \frac{1}{\gamma\,\ln(\gamma_2/\gamma_1)}&\alpha=1\\
 & \\
\frac{\alpha-1}{\gamma^{\alpha}}
 \left[\frac{(\gamma_1\gamma_2)^{\alpha-1}}
 {\gamma_2^{\alpha-1}-\gamma_1^{\alpha-1}}\right]&\alpha \neq1
\end{array}
\right.
\end{align}
We assume colored noise with $\alpha <2$ and, in particular, 
focus attention on classical noise with exponent in the range 
$\alpha\in[1/2,2]$. The case $\alpha=1$ is usually referred to as {\em pink
noise} and the case $\alpha=2$ as {\em Brown(ian) noise}.
\par
For colored noise the field $c(t)$ in Eq. \eqref{hamiltonian} is a 
superposition of $N$ random bistable fluctuators $c(t)=\sum_{j=1}^N c_j(t)$, 
where the $c_j(t)$ are classical stochastic fields describing independent
RTN sources with random switching rates extracted from the distribution \eqref{distrib}. 
The density matrix of a qubit interacting with colored noise is obtained as the 
average over all the environmental degrees of freedom \cite{benedettipra}:
\begin{align}
&\rho(\tau,\alpha,\theta,\phi)=\int_{\gamma_1}^{\gamma_2}\!\!
\rho(\tau,\gamma,\theta,\phi)\;p_{\alpha}(\gamma)\;d\gamma
\label{rcolor}
\end{align}
where $\rho(\tau,\gamma,\theta,\phi)$ is the expression of Eq. \eqref{rrtn} 
with the average taken over the global field $c(t)$. 
Eq. \eqref{rcolor} my be re-written as:
\begin{align}
\rho(\tau,\alpha,&\theta,\phi)=\nonumber\\
&\frac{1}{2}
(\cos\theta\, \mathbb{I}+ \Lambda(\tau,\alpha,N) \,\sigma_z+\nonumber\\
&\sin\theta\,\cos\phi\,\sigma_x-\sin\theta\,\sin\phi\,\Lambda(\tau,\alpha,N)\,\sigma_y)
\end{align}
where $N$ is the number of fluctuators, 
$\Lambda(\tau,\alpha,N)= \left[\Lambda(\tau,\alpha)\right]^N$, and 
\begin{equation}
 \Lambda(\tau,\alpha)=\int_{\gamma_1}^{\gamma_2}\!\! p_{\alpha}(\gamma)
 D(\tau,\gamma) d\gamma\label{lambda}\,. 
\end{equation}
The dynamics of the qubit is governed by the function $\Lambda$, which
can be easily evaluated numerically, either by numerical integration of 
Eq. (\ref{lambda}) or by the equivalent series representation reported in 
Appendix \ref{a:LL}.
\section{Local quantum estimation theory}\label{sec2}
In this section we review the main tools of local QET. Let us consider a family of 
quantum states $\rho_{\lambda}$ depending on a parameter $\lambda$. We
are interested in inferring the value of the parameter and to this aim we 
perform repeated measurements on the system and then process the overall 
sample of outcomes.
An estimator $\bar\lambda=\bar\lambda(x_1,x_2\dots x_M)$ 
is a function of the outcomes $\{x_i\}$ and we denote by 
$V(\bar\lambda)$ the corresponding mean square error. The smaller is 
$V(\bar\lambda)$, the more precise the estimator is. In fact, there is a 
bound to the precision of any unbiased estimator, given by the Cramer-Rao (CR) 
inequality:
\begin{equation}
V(\bar\lambda)\geq\frac{1}{M\, F(\lambda)}\label{ccr}
\end{equation}
where $M$ is the number of measurements and $F(\lambda)$ 
is the Fisher information:
\begin{equation}
F(\lambda)=\int\!\! dx\, p(x|\lambda)\, \left[\partial_\lambda\log
p(x|\lambda)\right]^2, 
\end{equation}
where $p(x|\lambda)$ is the conditional probability of obtaining 
the outcome $x$ when the true value of the parameter is $\lambda$.
In the case of a qubit, we may for instance consider the population 
measurement. 
The Fisher is given by:
\begin{equation}
F(\lambda)=\frac{(\partial_{\lambda}\rho_{00})^2}{\rho_{00}}+
\frac{(\partial_{\lambda}\rho_{11})^2}{\rho_{11}}
\end{equation}
where $\rho_{ii}$ are the two diagonal elements of the density matrix
in the population basis.
In order to compute the ultimate bound to precision as posed by quantum
mechanics, the FI must be 
maximized over all possible measurements. Upon introducing the Symmetric
Logarithmic Derivative $L_{\lambda}$ as the operator which satisfies the 
relation:
\begin{equation}
 \frac{L_{\lambda}\rho_{\lambda}+\rho_{\lambda}L_{\lambda}}{2}=\partial_{\lambda}\rho_{\lambda}
\end{equation}
the quantum CR bound is found:
\begin{equation}
 V(\lambda)\geq\frac{1}{M\, H(\lambda)}.\label{qcr}
\end{equation}
Here $H(\lambda)=\text{Tr}[\rho_{\lambda}L_{\lambda}^2]$
is the so-called quantum Fisher information.
In the case of a qubit, the expression of the QFI can be found
after diagonalizing the density matrix 
$\rho_{\lambda}=\sum_{n=1}^2 \rho_n|\phi_n\rangle\langle\phi_n|$:
\begin{equation}
H(\lambda)=\sum_{n=1}^2 \frac{(\partial_{\lambda}\rho_n)^2}{\rho_n}
+2\sum_{n\neq m}\frac{(\rho_n-\rho_m)^2}
{\rho_n+\rho_m}
 |\langle\phi_m|\partial_{\lambda}\phi_n\rangle|^2.\label{qfi}
\end{equation}
The first term in Eq. \eqref{qfi} is the classical FI of the distribution $\{\rho_n\}$,
while the second term has a quantum nature and vanishes when the 
eigenvectors of $\rho_{\lambda}$ do not depend upon the parameter $\lambda$.
When the condition $F(\lambda)=H(\lambda)$  is fulfilled, the 
measurement is said to be optimal. If equality in Eq. \eqref{ccr}
is satisfied the corresponding estimator is said to be efficient.
\par
A global measure of the estimability of a parameter is given by the
single-measurement signal-to-noise ratio $\snr=\lambda^2/V(\lambda)$. 
Using the Cramer-Rao bound  we have that the SNR is bounded by the 
so-called quantum signal-to-noise 
ratio $\qsnr$ $R=\lambda^2 H(\lambda)$, which 
represents the ultimate quantum bound
to the estimability of a parameter.
\section{Parameter estimation by quantum probes}\label{sec3}
The goal of an estimation procedure is not only to determine the 
value of an unknown parameter, but also to infer this value with the 
largest possible precision. The quantum CR inequality set a bound to 
the ultimate precision that can be achieved in estimating a 
parameter and, in turn, on the corresponding signal-to-noise ratio.
\par
In this section we discuss optimization of parameter estimation by
quantum probes. In other words, we determine the initial qubit 
preparation and the interaction time that maximize the QFI, and show 
that the corresponding ultimate precision may be achieved by population
measurement on the qubit. We then discuss in detail under which conditions 
it is possible to estimate efficiently the spectral properties of the 
environmental noise.
\par
Let us start by considering a generic pure dephasing model 
\begin{align}
 \rho=\frac{1+\Gamma(\lambda)}{2}\rho_0+\frac{1-\Gamma(\lambda)}{2}\sigma_x\rho_0\sigma_x
 \label{dephasing}
\end{align}
where $\Gamma(\lambda)$ is a real coefficient taking both negative and
positive values between $\pm 1$. If we set $\phi=0$ in Eq. 
\eqref{psi0}, the QFI can be
analytically computed and it takes the expression:
\begin{align}
 H(\lambda)=\cos^2(\theta)\;\frac{\big[\partial_{\lambda}\Gamma(\lambda)\big]^2}
 {1-\Gamma^2(\lambda)}.\label{qft}
\end{align}
As it is apparent from Eq. \eqref{qft} the QFI is maximized 
for $\theta=0$. In this case the optimal initial state preparation
is the state $\ket{\psi}=\ket{0}$.
If we consider the most general initial state \eqref{psi0} with $\phi\neq0$,
we have numerical evidence that the QFI is still maximized  by the 
state $|0\rangle$ for any choice of $\Gamma(\lambda)$.
\subsection{Random telegraph noise}
\begin{figure}[h]
\centering
\includegraphics[width=0.9\columnwidth]{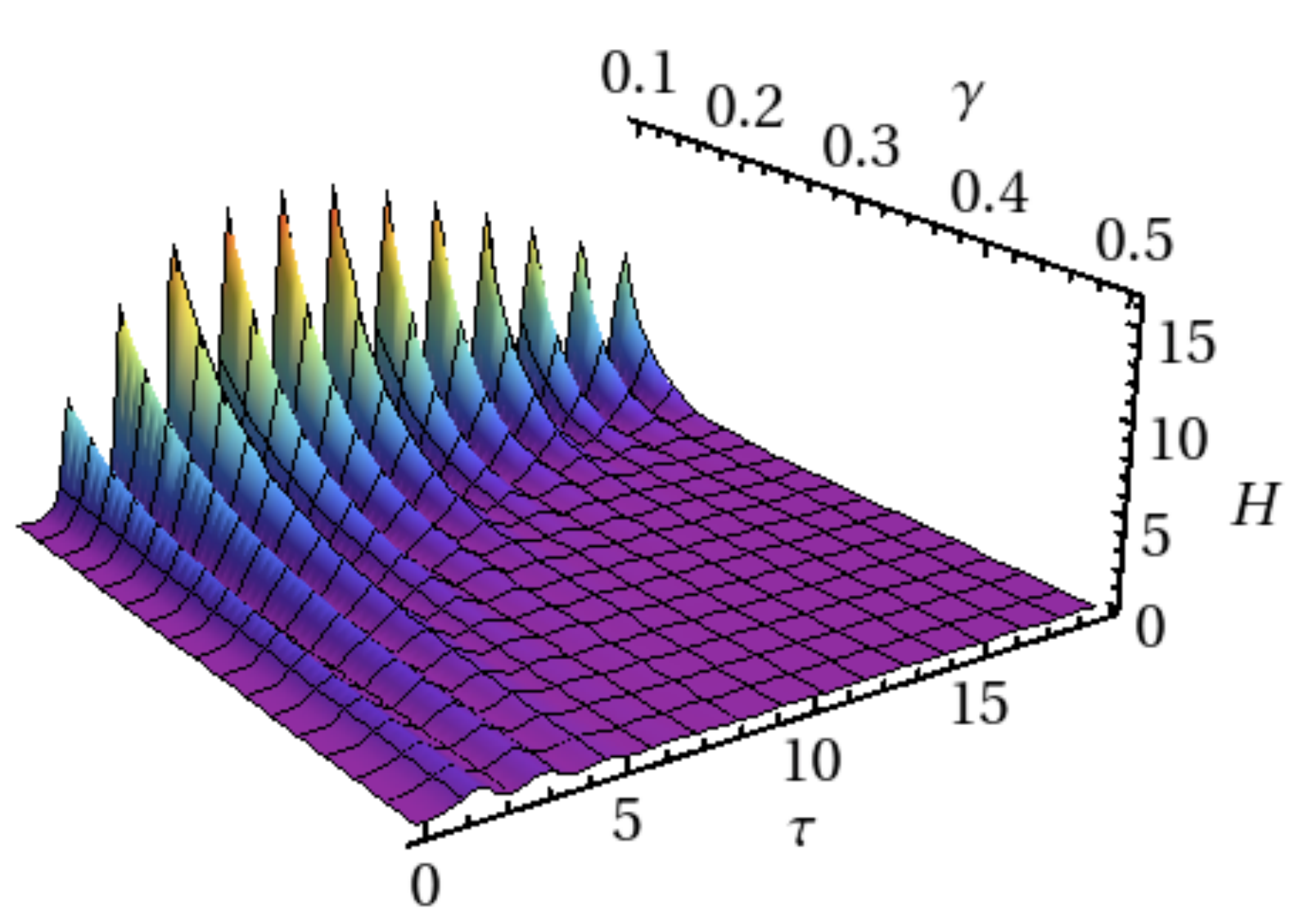}
\includegraphics[width=0.85\columnwidth]{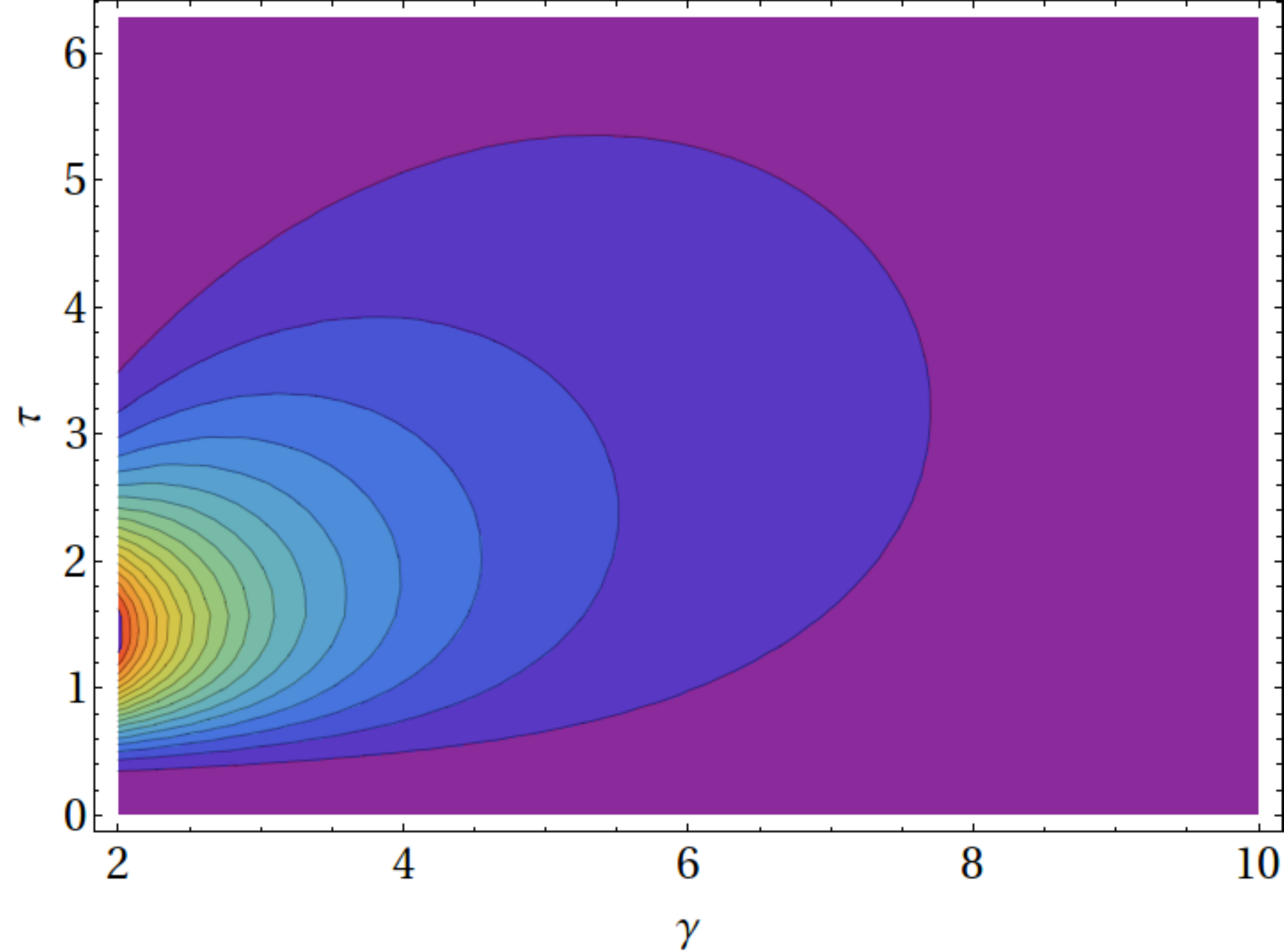}
\caption{(Color online): The upper panel shows the QFI $H(\tau,\gamma)$ 
as a function of the interaction time $\tau$ and the switching rate $\gamma$ 
for slow RTN. The lower panel shows a contour plot of of $H(\tau,\gamma)$ 
for fast RTN.}\label{fig1}
\end{figure}

In the case of a RTN the parameter to be estimated is the switching rate $\gamma$.
Starting from the qubit prepared in the state $|0\rangle$ and using Eq.
(\ref{rtntp}), the family of possible evolved states may be written as
\begin{align}
\rho(\tau,\gamma)=\frac{1}{2}\left(
\begin{array}{cc}
1+D(\tau,\gamma) & 0\\
0 &1-D(\tau,\gamma) \\\end{array}\right).\label{qrtn}
\end{align}
We know that the optimal measurement is a projective one
\cite{barn00,luati}. Besides, the eigenvectors of the matrix \eqref{qrtn} do not 
depend on the parameter $\gamma$ and 
the second term in Eq. \eqref{qfi} vanishes.
Looking at the very form of the matrices in Eq. \eqref{qrtn}
one immediately recognizes that the QFI coincides with 
the FI of population measurement and can be written as:
\begin{equation}
 H(\tau,\gamma)=\frac{[\partial_{\gamma}D(\tau,\gamma)]^2}{1-D(\tau,\gamma)^2},
 \end{equation}
which is the analogue of Eq. \eqref{qft} with the coefficient $\Gamma(\lambda)$ 
replaced by coefficient $D(\tau,\gamma)$. 
The two different regimes of slow and fast RTN give rise to different 
behaviors for the QFI, which are illustrated in Fig. \ref{fig1}. For slow 
RTN $H$ is shown in the upper panel of Fig. \ref{fig1}: the QFI 
is characterized  by an oscillating behavior and, in particular, for $\gamma\ll2$
the peaks are located at multiples of $\tau=\frac{\pi}{2}$. In the fast RTN case
(see the lower panel of Fig. \ref{fig1}), $H$ has
only one peak and its maximum value decreases with $\gamma$.
\par
\begin{figure}[t]
\centering
\includegraphics[width=0.9\columnwidth]{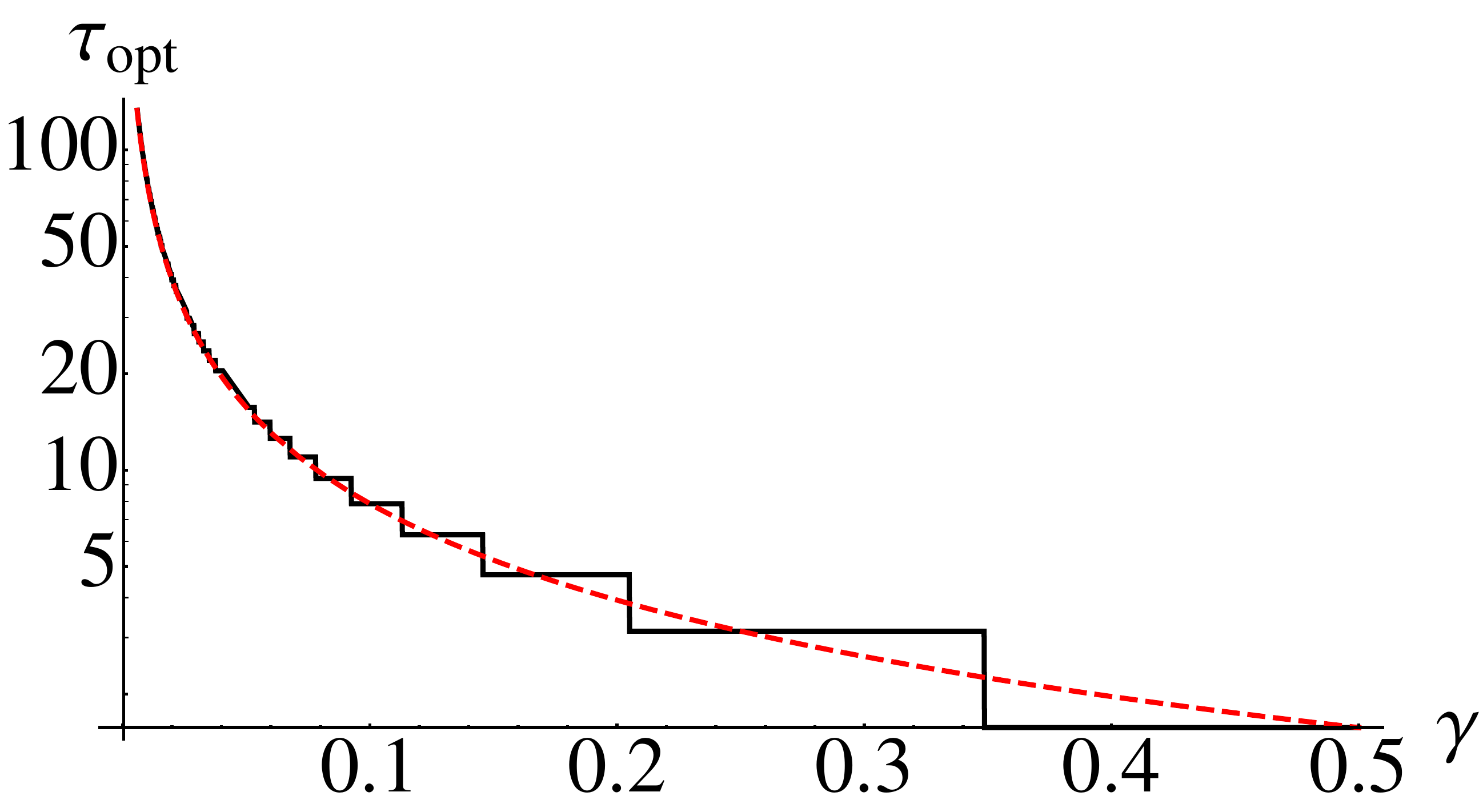}         
\caption{(Color online): The optimal interaction time $\tau_{\opt}$, maximizing 
the quantum Fisher information $H(\tau,\gamma)$ for slow RTN, as a function of the
switching rate $\gamma$ (black line). The dashed red curve denotes the function 
$\pi/4\gamma$.}\label{fig2}
\end{figure}

In order to optimize the inference procedure we look for the 
interaction time that maximizes the QFI $H(\tau,\gamma)$ (and, in turn,
the QSNR $R$) at each fixed value of the switching rate $\gamma$. The maximization of the QFI 
has been performed numerically, leading to the following approximation 
\begin{equation}
\tau_{\opt}(\gamma)\simeq \left\{\begin{array}{cc}
\text{nint}\left[\frac{1}{2\gamma}\right]\frac{\pi}{2}&\quad
\gamma<2\\
 &\\ \frac{2}{5}\gamma&\quad\gamma>2
\end{array}.
\right.\label{topt}
\end{equation}
The approximation is very good for $\gamma$ in range [$10^{-3},10^3$] 
except for $\gamma\simeq2$ where the peaks are not exactly located 
at multiples of $\tau=\frac{\pi}{2}$ and Eq. (\ref{topt}) is valid only
to a first approximation. In order to further illustrate the behavior
of the QFI in the slow RTN regime, in Fig. \ref{fig2} we show 
the optimal interaction time 
$\tau_{\opt}$ as a function of the switching rate. 
The step-like behavior of $\tau_{\opt}$ is due to 
the oscillating behavior of the QFI. On the other hand, in the fast RTN
regime, the maximum moves continuously as a function of $\gamma$.
\par
As seen from Eq. \eqref{topt}, optimal times increase with decreasing $\gamma$ in the
slow RTN regime and with increasing $\gamma$ in the fast RTN regime.
When small switching rates are considered, long times are necessary 
to see the effect of the environment on the 
probe, in agreement with the non Markovian character of the
corresponding evolution map \cite{Sun10,bpm13}. 
In the case $\gamma\gg2$, the qubit and 
the external fluctuators act as if they were decoupled, so
long observation times are required to see the influence of the 
external noise on the dynamics of the qubit.
In both cases, the maximum values $H(\tau_{\opt},\gamma)$ of the QFI 
are inversely proportional to $\gamma^2$. In particular, a numerical 
fit in range [$10^{-3},10^3$] leads to
\begin{equation}
H(\gamma)\approx \frac{a}{\gamma^2}.
\end{equation}
where $a$ is of the order of 0.1. The quantum signal-to-noise ratio
$\qsnr=\gamma^2 H(\gamma)\simeq a$ is thus constant, meaning that quantum probes 
allow one for a uniform estimation of the switching rate in the 
whole range of values we have considered.
\subsection{Colored noise}
In the case of a collection of random bistable fluctuators, the 
relevant parameter to be estimated is the ``color'' of the noise, 
i.e. the exponent $\alpha$. 
Following the general arguments mentioned at the beginning of this
Section we assume that the probe qubit is initially prepared in the 
state $\ket{0}$. Its time evolution is thus described by the 
density matrix:
\begin{align}
\rho(\tau,\alpha,N)=\frac{1}{2}\left(
 \begin{array}{cc}
 1+\Lambda(\tau,\alpha,N) & 0\\
 0 &1-\Lambda(\tau,\alpha,N) \\
\end{array}
\right).
\label{qalpha}
\end{align}
Also for colored noise 
the eigenvectors do not depend on the parameter $\alpha$
and thus the FI for population measurement coincides 
with the QFI, which is given by 
\begin{align}
H(\tau,\alpha,N)=N^2\;\frac{ \Lambda(\tau,\alpha)^{2N-2}\,}
{1-\Lambda(\tau,\alpha)^{2N}}
\Big[\partial_\alpha \Lambda(t,{\alpha})
\Big]^2\,.
\end{align}
For colored environment realized by a single random fluctuator 
the above formula reduces to
\begin{align}
H(\tau,\alpha)=\frac{\big[\partial_{\alpha} 
\Lambda(\tau,\alpha)\big]^2}{1-\Lambda(\tau,\alpha)^{2}}\,.
\end{align}
\begin{figure}[h]
\centering
\includegraphics[width=0.49\columnwidth]{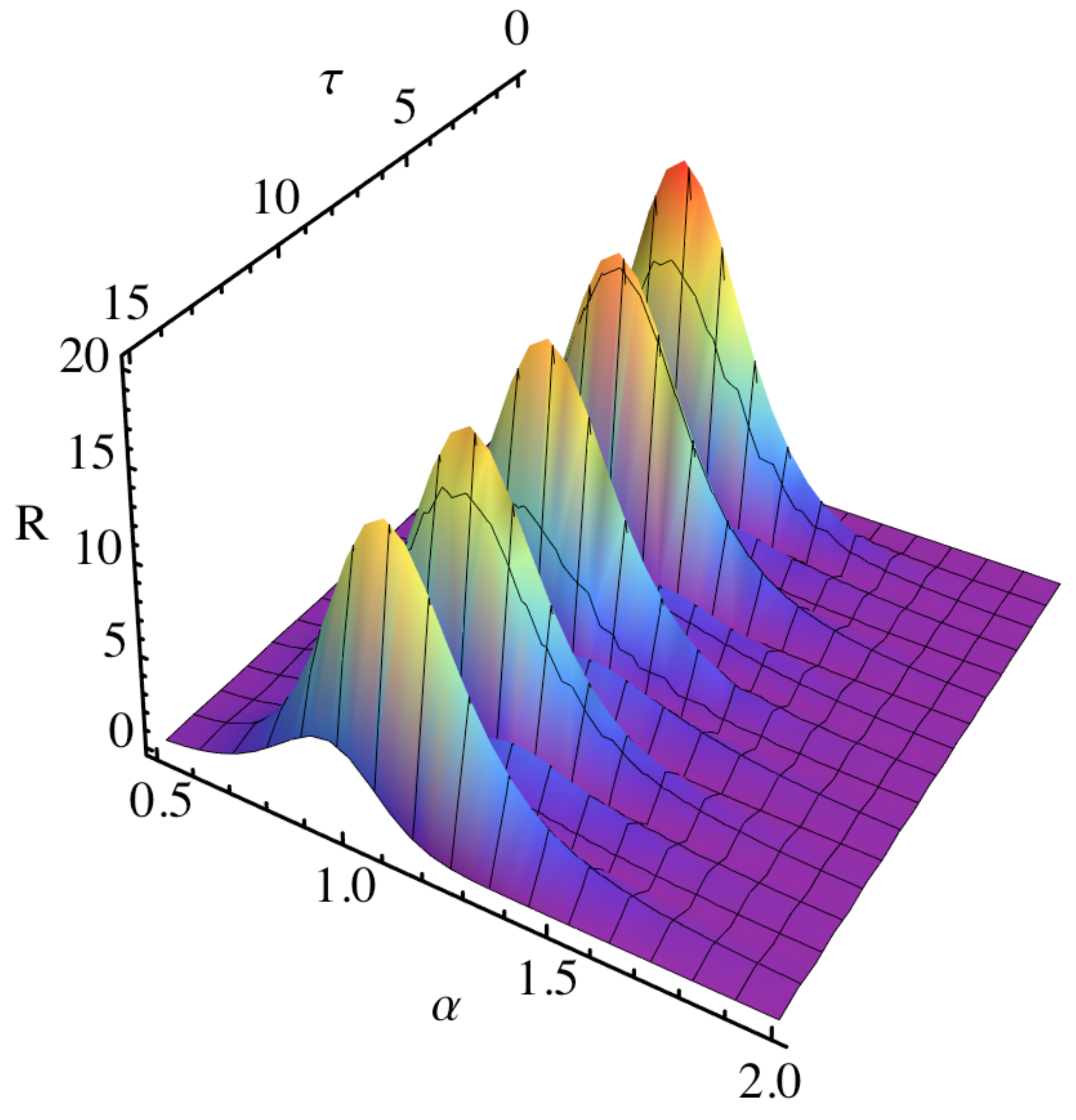}
\includegraphics[width=0.49\columnwidth]{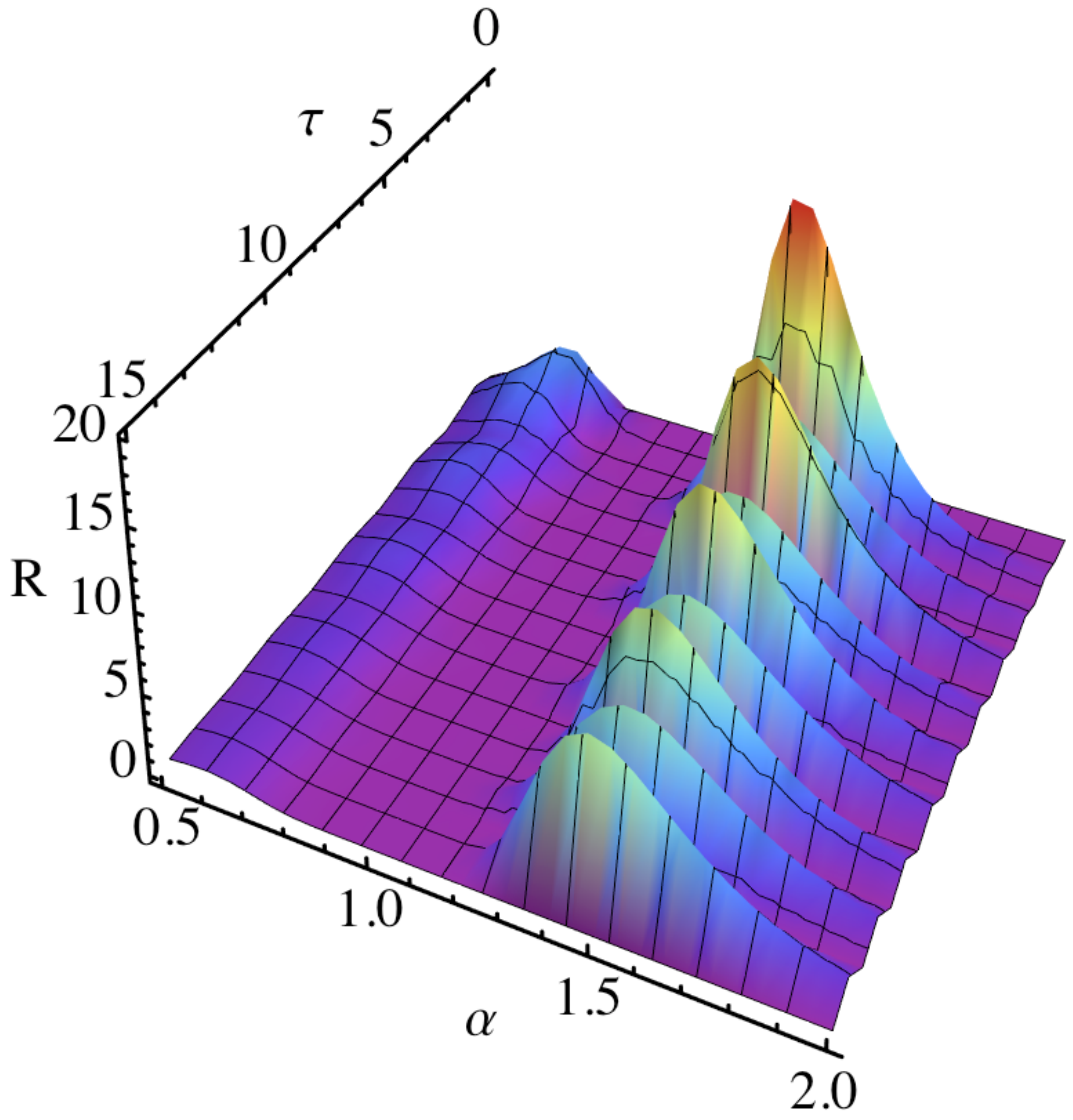}
\caption{(Color online) Spectral characterization of colored noise: The
left panel shows the QSNR $R(\tau,\alpha,N)$ as a function of the
interaction time and the exponent  $\alpha$ for a single fluctuator $N=1$. 
The right panel shows the same quantity for $N=10$.}  \label{fig3}
\end{figure}
\par
The QFI depends on the interaction time $\tau$, the exponent  $\alpha$
and the number of fluctuators $N$. Different values for 
$\alpha$ and $N$ may lead to considerably
different temporal behaviors for the QFI. This is illustrated  
in Fig. \ref{fig3}, where we show the QSNR $R(\tau,\alpha,N)=\alpha^2 H(\tau,\alpha,N)
$ as a function of $\alpha$ and $\tau$ for two different numbers of fluctuators.
When a single fluctuator is considered, the QSNR has a maximum located at 
$\alpha=1$, which corresponds to the best estimable value for 
the parameter. The situation is totally
reversed in the case of $N=10$ fluctuators, where values of $\alpha$
close to one correspond to a very low QSNR.
\par
\begin{figure}[h]
 \centering
 \includegraphics*[width=0.85\columnwidth]{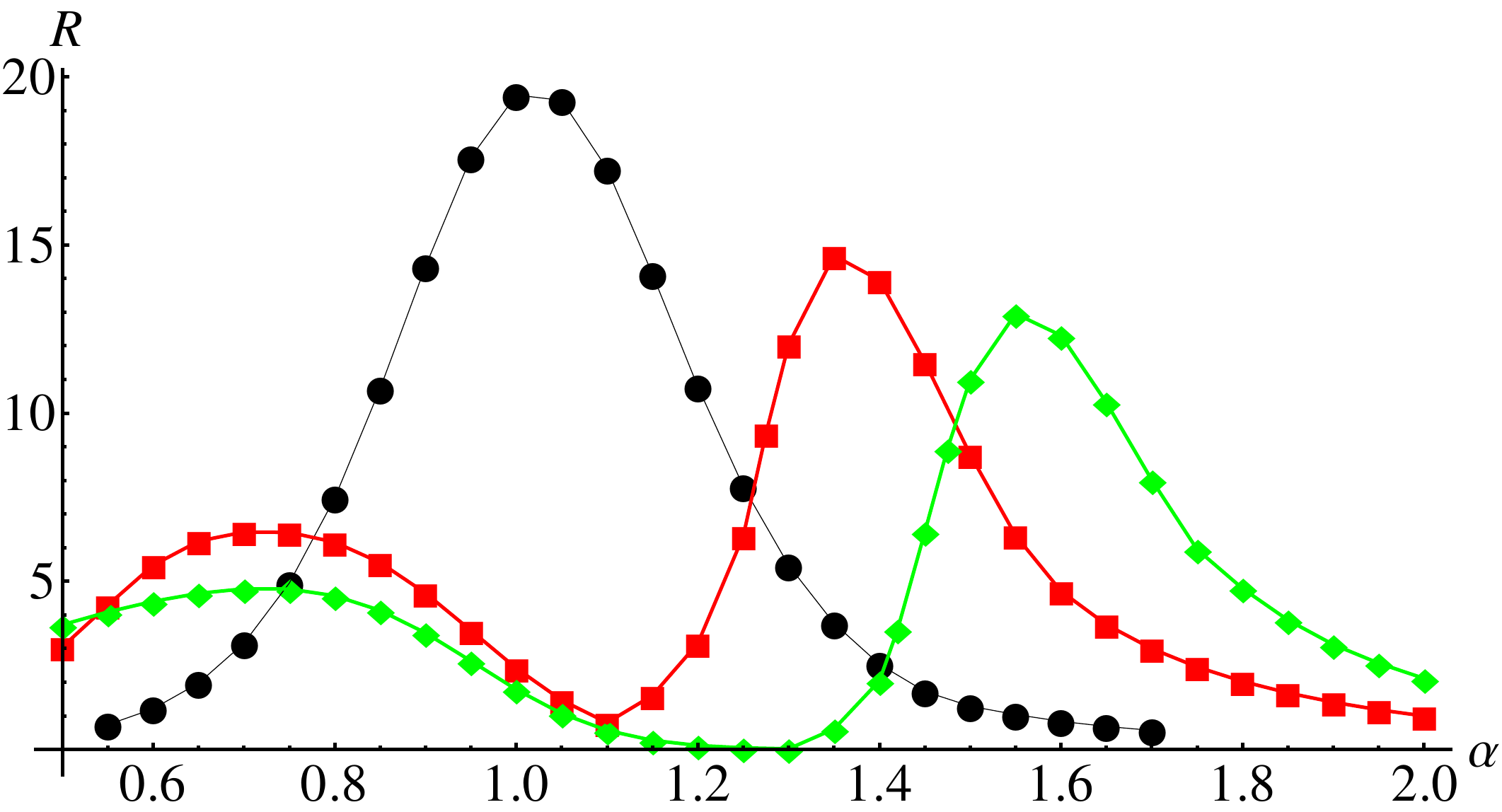}
\caption{(Color online) Spectral characterization of colored noise: The
plot shows the QSNR $R(\tau_{\opt},\alpha,N)$ as a function of $\alpha$ 
for different numbers of fluctuators: $N=1$ (black circles), 
$N=10$ (red squares) and $N=50$ (green rhombuses). Lines are guides 
for the eyes.}  \label{fig4}
\end{figure}

In order to further illustrate this behavior, 
in Fig. \ref{fig4} we show the QSNR, already maximized over the
interaction time,  as a function of 
$\alpha$ for (three) fixed numbers of fluctuators.
For a single fluctuator the QSNR exhibits a single
maximum located at $\alpha=1$, i.e. pink noise is
more precisely estimable than other kind of noise.
On the other hand, when the number of fluctuators
increases, two maxima appear and their location move
away from $\alpha=1$ for increasing $N$, with the
largest maximum drifting towards $\alpha=2$.
\par
\begin{figure}[h!]
 \centering
 \includegraphics*[width=0.89\columnwidth]{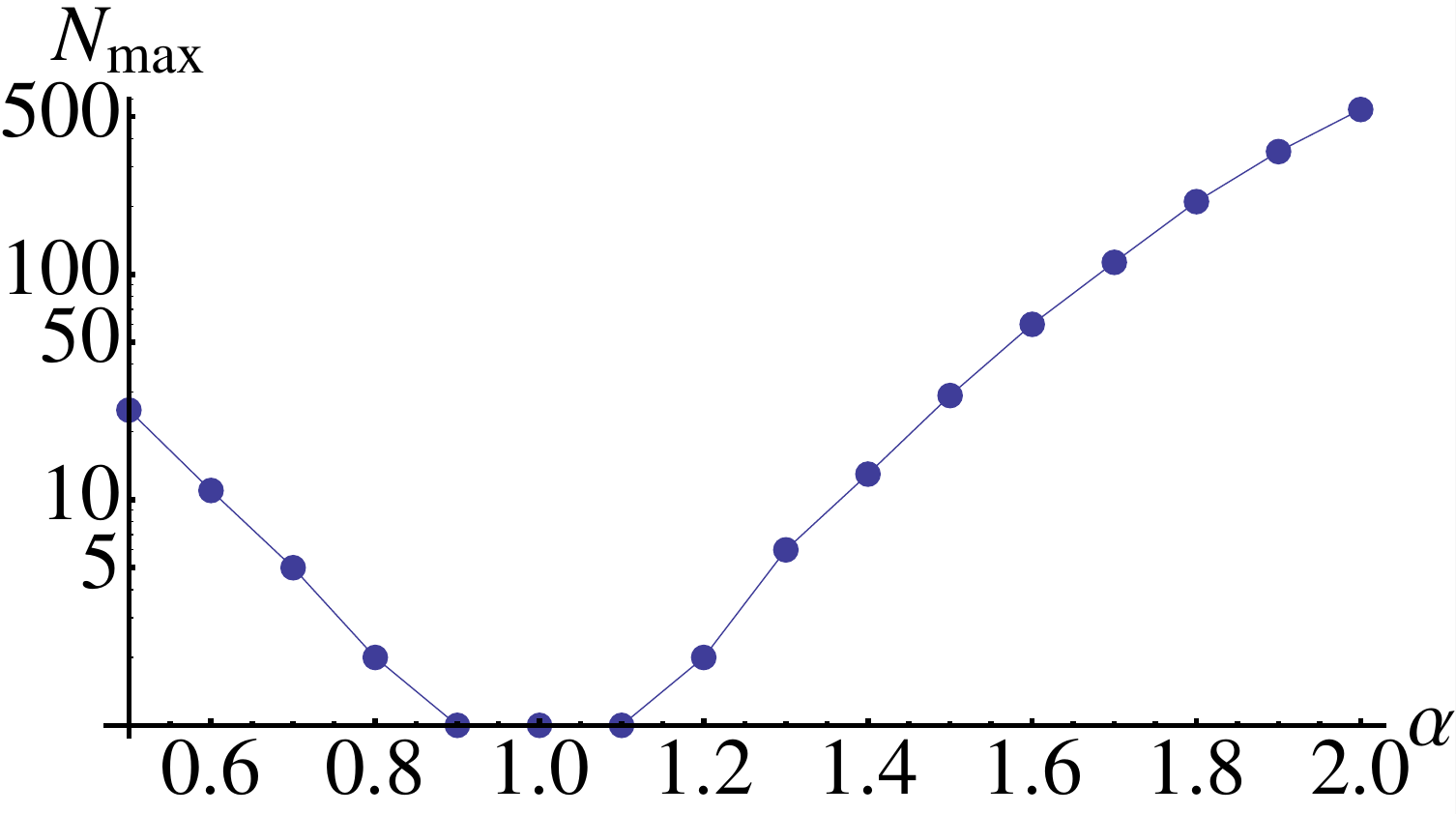}
\caption{(Color online) Spectral characterization of colored noise: The
plots shows the number of fluctuators $N_{\max}$ that maximizes the QFI
as a function of the exponent $\alpha$. The line is a guide for the eyes.} 
\label{fig5}
\end{figure}

To complete our analysis we also investigate with some more details 
the dependence of the QFI on the structure of the environment, i.e. 
on the number of fluctuators describing the environment.
In Fig. \ref{fig5} we show the number of fluctuators $N_{\max}$ maximizing
the QFI as a function of $\alpha$. We first notice that
there is indeed a dependence, and that $N_{\max}$ may be considerably
different for, say, pink or brown noise. As it is apparent from
Fig. \ref{fig5} $N_{\max}$ decreases with increasing $\alpha$ until it reaches
the value $N_{\max}=1$ for values of $\alpha$ close to 1. Then
it increases with $\alpha$, up to $N_{\max}=540$ 
for $\alpha=2$. \par 
As a final remark, we also notice that when the number of fluctuators 
is taken equal to $N_{\max}$, than the optimal interaction time maximizing
the QFI is $\tau\simeq\pi/2$ independently on $\alpha$.
\section{Conclusions}\label{sec4}
In this paper we have addressed the estimation of the spectral properties 
of classical environments using a qubit as a quantum probe. In particular, 
we have focused attention on the estimation of the switching rate
$\gamma$
of random telegraph noise and of the exponent of colored noise 
with $1/f^{\alpha}$ spectrum.  In both cases we have evaluated 
the quantum Fisher information and found the optimal initial 
preparation and the optimal interaction time that maximize its value.
We have also shown that 
population measurement on the qubit is optimal, that is 
the Fisher information coincides with the quantum Fisher information.  
\par
For random telegraph noise the (maximized) QFI is inversely proportional 
to the square of the switching rate, meaning that the quantum signal-to-noise ratio
is constant and thus the switching rate may be estimated with uniform
precision 
in its whole range of variation. The corresponding value
of the optimal interaction time decreases with increasing $\gamma$ and is located 
at multiple of $\pi/2$ in the slow RTN regimes, whereas it grow linearly 
with $\gamma$ in the fast RTN regime.
\par
For colored noise, we studied the estimability of the color of the
spectrum, i.e. of the exponent $\alpha$. Our results show that two
different cases emerges: if the environment is modeled by a single
random fluctuator, then estimation is more precise for pink noise, i.e.
for $\alpha=1$.  On the other hand, when the environment is instead
described as a collection of several fluctuators, the QFI is has two
local maxima, whose positions drift towards the boundaries of the
interval $[0.5,2]$ as $N$ is increased. The largest quantum
signal-to-noise ratio is obtained for brown noise, i.e. for
$\alpha\simeq 2$.
We also find that for any fixed value of $\alpha$ there is
a specific number of fluctuators maximizing the QFI for the interaction 
time $\tau\simeq\pi/2$, independently on the value of $\alpha$.
\par
Overall, our results show that the features of a complex environment may be
reliably determined by monitoring a small quantum probe with 
more easily controllable degrees of freedom.  In
particular, our results show that quantum probes permit to reliably
estimate the characteristic parameters of classical noise using
measurements performed after a fixed optimal interaction time, rather
than collecting a series of measurements to estimate the autocorrelation
function of the underlying stochastic process.  
\acknowledgments
This work has been supported by the MIUR project FIRB-LiCHIS-RBFR10YQ3H. 
\appendix
\section{Series representation for $\Lambda(\tau,\alpha)$}
\label{a:LL}
Using the expression (\ref{distrib}) of the distribution
$p_{\alpha}(\gamma)$ 
Eq. (\ref{lambda}) may be rewritten as 
\begin{align}
\Lambda(\tau,\alpha) = & N_{\alpha} (\gamma_1,\gamma_2) 
\int^{\gamma_2}_{\gamma_1}\!\! d\gamma\, e^{-\gamma \tau}
\gamma^{-\alpha}  
\notag \\
& \times 
\Big[ \cosh \delta\tau 
+ \gamma\tau \frac{\sinh\delta\tau}{\delta\tau}
\Big] \label{a1}
\end{align}
where $\delta=\sqrt{\gamma^2-4}$  and the normalization reads as follows
\begin{align}\label{norms}
N_{\alpha}(\gamma_1,\gamma_2)=\left\{\begin{array}{ll}
 \frac{1}{\ln\gamma_2-\ln\gamma_1}&\alpha=1\\
 & \\
(\alpha-1)
 \left[\frac{(\gamma_1\gamma_2)^{\alpha-1}}
 {\gamma_2^{\alpha-1}-\gamma_1^{\alpha-1}}\right]&\alpha \neq1
\end{array}
\right.\,.
\end{align}
Using the new variable $y=\gamma\tau$ we may write 
\begin{align}
\Lambda(\tau,\alpha) =  N_{\alpha} (\gamma_1, \gamma_2) \Big[
F(\gamma_2\tau,\alpha,\tau) - F(\gamma_1\tau,\alpha,\tau)
\Big]\,,\label{a3}
\end{align}
where 
$$F(y,\alpha,\tau)=\tau^{\alpha-1} \Big[F_1(y,\alpha,\tau) +
F_2(y,\alpha,\tau)\Big]\,,$$
and
\begin{align}
F_1(y,\alpha,\tau) &= 
\int\!\! dy\, e^{-y}\, y^{-\alpha}\, \cosh \sqrt{y^2-4\tau^2}\,, \\  
F_2(y,\alpha,\tau) &=
\int\!\! dy\, e^{-y}\, y^{-\alpha+1}\, \frac{\sinh
\sqrt{y^2-4\tau^2}}{\sqrt{y^2-4\tau^2}}\,.
\end{align}
Upon expanding the hyperbolic functions and using the relation
\begin{align}
\int\!\! dy\, e^{-y}\, y^{-\alpha}\, (y^2-4\tau^2)^k
=& \sum_{p=0}^k\, (-)^{1+k+p}\, (2\tau)^{2(k-p)} 
\notag \\ & \times 
\left(\begin{array}{c} k \\ p \end{array}\right)
\Gamma(2p+1-\alpha,y)\,,
\end{align}
where $\Gamma(a,x)$ is the (incomplete) Euler Gamma function, 
the two functions $F_k$ may be rewritten as
\begin{align}
F_1(y,\alpha,\tau) = &
\sum_{k=0}^\infty \sum_{p=0}^k\,(-)^{1+k+p}\, \frac{\tau^{2(k-p)}}{(2k)!}
\notag \\ &\times \left(\begin{array}{c} k \\ p \end{array}\right)
\Gamma(2p+1-\alpha,y)\,, \\ 
F_2(y,\alpha,\tau) = &
\sum_{k=0}^\infty \sum_{p=0}^k\,(-)^{1+k+p}\, \frac{\tau^{2(k-p)}}{(2k+1)!}
\notag \\ &\times \left(\begin{array}{c} k \\ p \end{array}\right)
\Gamma(2p+2-\alpha,y)\,.
\end{align}
We now introduce the new index $s=k-p$ and rearrange series as
$$
\sum_{k=0}^\infty \sum_{p=0}^k ... = 
\sum_{p=0}^\infty \sum_{k=p}^\infty ... = 
\sum_{p=0}^\infty \sum_{s=0}^\infty ...\,,$$
thus arriving at
\begin{align}
F_1(y,\alpha,\tau) = &
\sum_{p=0}^\infty \sum_{s=0}^\infty \frac{(-)^{1+s}}{[2(p+s)]!} 
\left(\begin{array}{c} p+s \\ s \end{array}\right) (2\tau)^s
\notag \\ & \times \Gamma(2p+1-\alpha,y) \nonumber\\
= & - \sum_{p=0}^\infty \frac{1}{(2p)!}\,
\Phi_{p+\frac12} (-\tau^2)\,\Gamma(2p+1-\alpha,y)\,,\label{a9}
\end{align}
where $\Phi_n(x)$ denotes the confluent hypergeometric function
${}_0F_1(n,x)$. Analogously, we arrive at
\begin{align}
F_2(y,\alpha,\tau) = & - \sum_{p=0}^\infty \frac{1}{(2p+1)!}\,
\Phi_{p+\frac32} (-\tau^2)\,\Gamma(2p+2-\alpha,y)\,.\label{a10}
\end{align}
Upon substituting Eqs. \eqref{a9} and \eqref{a10} in Eq. \eqref{a3} we obtain
a series representation for the quantity $\Lambda(\tau, \alpha)$. As a matter of
fact, truncating the series at the first term, i.e. $p = 0$ in Eqs.
 \eqref{a9} and \eqref{a10}, already provides an excellent approximation
for $\alpha\gtrsim 3/2$ and any value of $\tau$. In formula
\begin{align}
 \Lambda(\tau, \alpha)\simeq&\frac{1}{2}N_{\alpha}(\gamma_1 , \gamma_2 )\tau^{ \alpha-2} \big[2\tau
 \cos 2\tau \,\Gamma(1 - \alpha, \gamma_1 \tau, \gamma_2 \tau )\nonumber\\
 &+ \sin 2\tau\, \Gamma(2 - \alpha, \gamma_1 \tau, \gamma_2 \tau ) \big],
\end{align}
where $\Gamma(a, x, y) = \Gamma(a, x) -\Gamma(a, y)$. On the other hand, for
$\alpha\lesssim 3/2$ the number of terms needed for a reliable approximation rapidly grows.


\begin{thebibliography}{99}
\bibitem{band13} P. Szakowski, M. Trippenbach, and Y. B. Band, 
Phys. Rev. E {\bf 87}, 052112 (2013).
\bibitem{berman12} A. I. Nesterov, G. P. Berman,  Phys. Rev. A {\bf 85}, 052125 (2012). 
\bibitem{galp06} Y. M. Galperin, B. L. Altshuler, J. Bergli, and D. V. Shantsev, 
Phys. Rev. Lett. {\bf 96}, 097009 (2006). 
\bibitem{joynt13} D. Crow, R. Joynt,  arXiv:1309.6383v1.
\bibitem{helm11} J. Helm, W. T. Strunz, S. Rietzler, and L. E. Warflinger, 
Phys. Rev. A {\bf 83}, 042103 (2011).
\bibitem{hur51} H. E. Hurst, Trans. Am. Soc. Civil. Eng. {\bf 116}, 770 (1951).
\bibitem{per03} D. B. Percival, Metrologia {\bf 40}, S289 (2003).
\bibitem{barunik10} J. Barunik, L. Kristoufek, Physica A {\bf 389}, 3844 (2010).
\bibitem{kend99} C. M. Kendziorski, J. B. Bassingthwaighte, P. J. Tonellato,
Physica A {\bf 273}, 439 (1999).
\bibitem{salomon13} L. A. Salomon, J. C. Fort, J. Stat. Comp. Simul. {\bf 83}, 542 (2013).
\bibitem{magd13}  M. Magdziarz, J. K. Slezak, J. Wojcik, J. Phys. A {\bf 46}, 325003
(2013).
\bibitem{galperin} J. Bergli, Y. M. Galperin and B. L. Altshuler, New J.
Phys. {\bf 11} 025002 (2009). 
\bibitem{benIJ} C. Benedetti, F. Buscemi, P. Bordone, M. G. A. Paris, 
Int. J. Quantum Inf. {\bf 10}, 1241005 (2012).
\bibitem{bordonefnl} P. Bordone, F. Buscemi, and 
C. Benedetti, Fluct. Noise Lett. {\bf 11}, 1242003 (2012).
\bibitem{weissman}M. B. Weissman Rev. Mod. Phys. {\bf 60} 537 (1998).
\bibitem{paladinorev} E. Paladino, Y. M. Galperin, G. Falci, B. L. Altshuler,  arXiv:1304.7925v1.
\bibitem{benedettipra} C. Benedetti, F. Buscemi, P. Bordone, M. G. A. Paris, Phys. Rev. A {\bf 87}, 052328 (2013).
\bibitem{bus13} F. Buscemi, P. Bordone, Phys. Rev. A {\bf 87}, 042310
(2013).
\bibitem{hel76} C. W. Helstrom, Quantum Detection and Estimation Theory
(Academic Press, New York, 1976).
\bibitem{mal93} J. D. Malley, J. Hornstein, Statist. Sci. {\bf 8}, 433 (1993).
\bibitem{brau94} S. Braunstein, C. Caves, Phys. Rev. Lett. {\bf 72}, 3439 (1994).
\bibitem{brody99} D. C. Brody, L. P. Hughston, Proc. Roy. Soc. Lond. A {\bf 454}, 2445
(1998); A {\bf 455}, 1683 (1999).
\bibitem{paris09} M. G. A. Paris, Int. J. Quant. Inf. {\bf 7}, 125 (2009).
\bibitem{esch12} B. M. Escher, L. Davidovich, N. Zagury, R. L. de Matos Filho,
Phys. Rev. Lett. {\bf 109}, 190404 (2012);
B. M. Escher, R. L. de Matos Filho, and L. Davidovich, Braz. J. Phys.
{\bf 41}, 229 (2011).
\bibitem{hotta05} M. Hotta, T. Karasawa, M. Ozawa, 
Phys. Rev. A {\bf 72}, 052334 (2005).
\bibitem{monras07} A. Monras, M. G. A. Paris Phys. Rev. Lett. {\bf 98}, 160401 (2007).
\bibitem{fuj01} A. Fujiwara, Phys. Rev. A {\bf 63}, 042304 (2001); A. Fujiwara, H.
Imai, J. Phys. A {\bf 36}, 8093 (2003).
\bibitem{ji08} Z. Ji, G. Wang, R. Duan, Y. Feng, M. Ying IEEE Trans. Inf.
Theory, {\bf 54}, 5172 (2008).
\bibitem{dau06} V. D’Auria, C. de Lisio A. Porzio, S. Solimeno, and M. G. A.
Paris J. Phys. B {\bf 39}, 1187 (2006).
\bibitem{brida2010} G. Brida, I. Degiovanni, A. Florio, M. Genovese, P. Giorda,
A. Meda, M. G. A. Paris, A. Shurupov, Phys. Rev. Lett. {\bf 104},
100501 (2010).
\bibitem{brida2011} G. Brida, I. P. Degiovanni, A. Florio, M. Genovese, P. Giorda,
A. Meda, M. G. A. Paris, and A. P. Shurupov, Phys. Rev. A {\bf 83},
052301 (2011) .
\bibitem{blandino2012} R. Blandino, M. G. Genoni, J. Etesse, M. Barbieri, M. G. A.
Paris, P. Grangier, and R. Tualle-Brouri, Phys. Rev. Lett. {\bf 109},
180402 (2012).
\bibitem{benedetti2013} C. Benedetti, A. P. Shurupov, M. G. A. Paris, G. Brida, and M.
Genovese, Phys. Rev. A {\bf 87}, 052136 (2013).
\bibitem{mon10} A. Monras, F. Illuminati Phys. Rev. A {\bf 81}, 062326 (2010); Phys.
Rev. A {\bf 83}, 012315 (2011).
\bibitem{pinel13} D. Braun, J. Martin, Nat. Comm. {\bf 2}, 223 (2011);
O. Pinel, P. Jian, N. Treps, C. Fabre, and D. Braun Phys. Rev.
A {\bf 88}, 040102(R) (2013)
\bibitem{monar} A. Monras, preprint ArXiv:1303.3682.
\bibitem{monras06} A. Monras, Phys. Rev. A {\bf 73}, 033821 (2006).
\bibitem{kac10} M. Kacprowicz, R. Demkowicz-Dobrzanski, W. Wasilewski, K.
Banaszek, I. A. Walmsley, Nature Phot. {\bf 4}, 357 (2010).
\bibitem{cable10} H. Cable, G. A. Durkin, Phys. Rev. Lett. {\bf 105}, 013603 (2010).
\bibitem{durk10} G. A. Durkin, New J. Phys. {\bf 12} 023010 (2010).
\bibitem{genoni2011} M. G. Genoni, S. Olivares, M. G. A. Paris, Phys. Rev. 
Lett. {\bf 106}, 153603 (2011);
M. G. Genoni, S. Olivares, D. Brivio, S.  Cialdi, D. Cipriani, A.
Santamato, S. Vezzoli, M. G. A. Paris, Phys. Rev. A {\bf 85}, 043817 (2012).
\bibitem{spagnolo2012} N. Spagnolo, C. Vitelli, V. G. Lucivero, 
V. Giovannetti, L. Maccone, and F. Sciarrino, 
Phys. Rev. Lett. {\bf 108}, 233602 (2012). 
\bibitem{zan08} P. Zanardi, M. G. A. Paris, L. Campos-Venuti, Phys. Rev. A {\bf 78},
042105 (2008).
\bibitem{inver08} C. Invernizzi, M. Korbmann, L. Campos-Venuti, M. G. A. Paris,
Phys. Rev. A {\bf 78}, 042106 (2008).
\bibitem{brunelli2011} M. Brunelli, S. Olivares, M. G. A. Paris, 
Phys. Rev. A {\bf 84}, 032105 (2011);
M. Brunelli, S. Olivares, M. Paternostro, M. G. A. Paris, 
Phys. Rev. A 86, 012125 (2012).
\bibitem{benedetti1} C. Benedetti, F. Buscemi, P. Bordone, M. G. A. Paris, Int. J. Quant. Inf. {\bf 10}, 1241005 (2012).
\bibitem{paladino}R Lo Franco, A D'Arrigo, G Falci, G Compagno and E Paladino, Physica Scripta {\bf T147}, 014019 (2012).
\bibitem{abel}B. Abel and F. Marquardt, Phys. Rev. B {\bf 78}, 201302(R) (2008). 
\bibitem{bpm13} C. Benedetti, M. G. A. Paris and S. Maniscalco, arXiv:1309.5270v2.
\bibitem{barn00} O. E. Barndorff-Nielsen, R. D. Gill, J. Phys. A {\bf 33}, 4481 (2000).
\bibitem{luati}A. Luati, Annals of Statistics {\bf 32}, 1770, (2004).
\bibitem{Sun10} X.-M. Lu, X. Wang, C. P. Sun, Phys. Rev. A {\bf 82},
042103 (2010).
\end{thebibliography}
\end{document}